\documentclass[preprint,superscriptaddress,showpacs,amsmath,amssymb,aps,pra]{revtex4-1}
\usepackage{times}
\usepackage{amssymb}
\usepackage{graphicx}
\usepackage{dcolumn}
\usepackage{dsfont}
\usepackage{bm}
\usepackage{subfigure}
\usepackage[ruled]{algorithm2e}
\usepackage{hyperref}
\usepackage{appendix}
\usepackage{enumerate}
\usepackage{amsmath}
\makeatletter
\renewcommand*\env@matrix[1][\arraystretch]{%
  \edef\arraystretch{#1}%
  \hskip -\arraycolsep
  \let\@ifnextchar\new@ifnextchar
  \array{*\c@MaxMatrixCols c}}
\makeatother
\newtheorem{theorem}{Theorem}[section]
\newtheorem{lemma}[theorem]{Lemma}
\newtheorem{definition}[theorem]{Definition}
\newtheorem{construction}[theorem]{Construction}

\newcommand{\p}{\ensuremath{{\bf Proof.\ }}}

\newcommand{\qed}{\hphantom{.}\hfill $\Box$\medbreak}

\hypersetup{colorlinks=true, citecolor=blue, urlcolor=blue, linkcolor=blue}

\begin{document}
\title{Quantum $k$-uniform states from quantum orthogonal arrays}
\author{Yajuan Zang}
\affiliation{School of Mathematical Sciences, Capital Normal University,  Beijing, 100048, China}
\author{Zihong Tian}
\email{tianzh68@163.com}
\affiliation{School of Mathematical Sciences, Hebei Normal University, Shijiazhuang, 050024, China}
\author{Shao-Ming Fei}
\affiliation{School of Mathematical Sciences, Capital Normal University, Beijing, 100048, China}
\author{Hui-Juan Zuo}
\affiliation{School of Mathematical Sciences, Hebei Normal University, Shijiazhuang, 050024, China}

\begin{abstract}
The quantum orthogonal arrays define remarkable classes of multipartite entangled states called $k$-uniform states whose every reductions to $k$ parties are maximally mixed. We present constructions of quantum orthogonal arrays of strength 2 with levels of prime power, as well as some constructions of strength 3. As a consequence, we give infinite classes of 2-uniform states of $N$ systems with dimension of prime power $d\geq 2$ for arbitrary $N\geq 5$; 3-uniform states of $N$-qubit systems for arbitrary $N\geq 6$ and $N\neq 7,8,9,11$; 3-uniform states of $N$ systems with dimension of prime power $d\geq 7$ for arbitrary $N\geq 7$.

\end{abstract}
\parskip=3pt

\maketitle

\section{Introduction}
\label{intro}
Quantum entanglement is considered to be one of the most striking features of quantum mechanics. It has been widely utilized as a crucial resource in quantum information science~\cite{Chuang,Benenti} such as quantum computation~\cite{Jozsa}, quantum teleportation~\cite{BennettC2} and quantum key distribution~\cite{BennettC1,Lo}. Recently, a special class of multipartite entangled states has attracted much attention for a wide range of quantum tasks. These states are called $k$-uniform states which have the property that all of their reductions to $k$ parties are maximally mixed \cite{Goyeneche0}. An $N$-qudit state $|\Phi\rangle$ in Hilbert space $\mathcal{H}(N,d):=\mathbb{C}_d^N$ is $k$-uniform whenever
\begin{equation}
\rho_S=Tr_{S^c}|\Phi\rangle\langle\Phi|\propto I,~~~~\forall S\subset\{1,2,\ldots, N\},~ |S|\leq k,
\end{equation}
where $I$ is the identity matrix and $S^c$ denotes the complementary set of $S$. The Schmidt decomposition implies that a state can be at most $\lfloor N/2\rfloor$-uniform,
i.e., $k\leq \lfloor N/2\rfloor$. In addition, a $k$-uniform state is still a $(k-1)$-uniform state. The $\lfloor N/2\rfloor$-uniform state is called absolutely maximally entangled state denoted by AME($N,d$). AME($N,d$) exhibits maximal entanglement in all possible partitions and thus plays a pivotal role in quantum secret sharing, multipartite teleportation and in tensor network states for holographic codes~\cite{Zhang,Latorre}.

A plenty of works has been done to find $k$-uniform states and their applications ~\cite{Facchi1,Facchi2,Helwig}. Many results are based on irredundant orthogonal array in combinatorial design. So far, 2-uniform states for any $d\geq 2$, $N\geq 4$ except for $d=2, N=4$ \cite{Scott,Goyeneche0,Li,Zang1,Higuchi,Pang1,Rather,Zang2}, and 3-uniform states for any $d\geq 2$, $N\geq 6$ except for $d\equiv 2$ (mod 4), $N=7$ \cite{Helwig,Huber1,Li,Zang,Pang1} have been obtained. Especially, AME(4,$d$) for $d\neq 2$, AME(5,$d$) for any $d$, AME(6,$d$) for any $d$ and  AME(7,$d$) for $d\not \equiv 2$ (mod 4) \cite{Huber1,Huber,Scott,Pang1,Li,Rains} have been presented. There are also some results on higher uniformity $k\geq 4$ \cite{Grassl,Pang2,Chen}. Incidentally, we notice that these states gained by orthogonal arrays satisfy a strong condition, i.e., they have the same coefficient in each items, which may lead to certain applications in quantum information processing.

\begin{figure}
\centering
\subfigure{
\includegraphics[width=10cm,height=4cm]{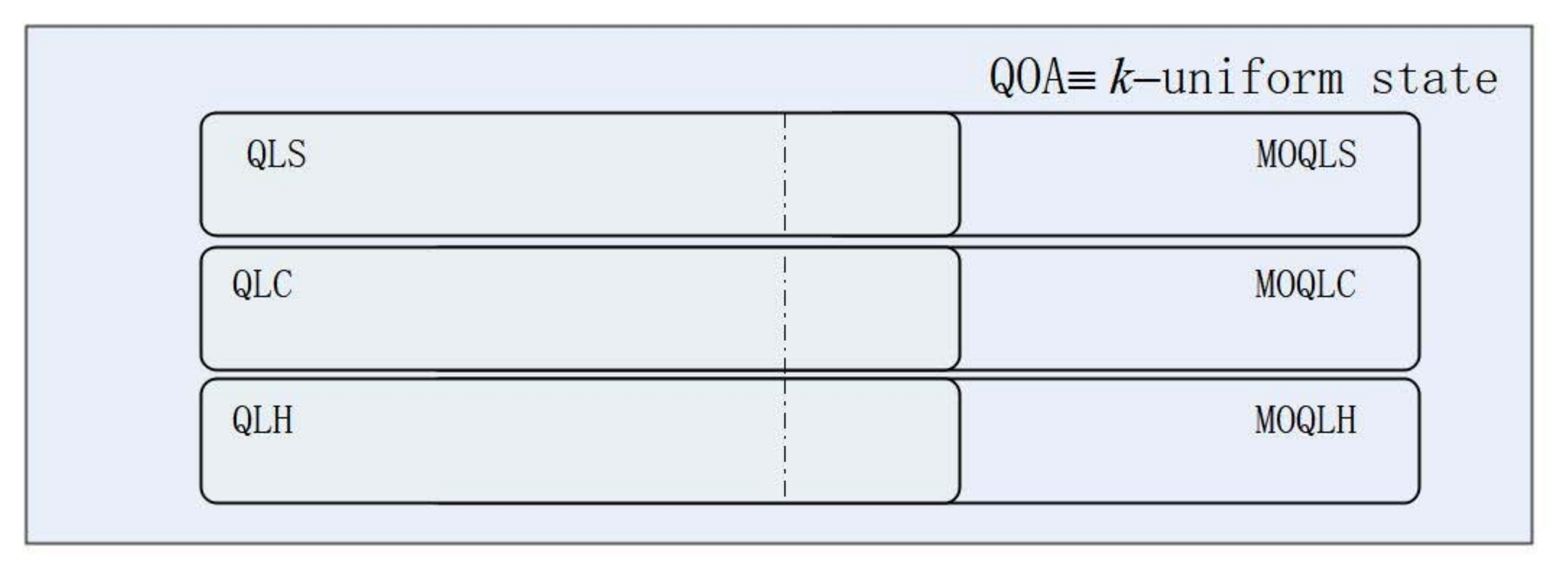}}
\caption{The relationships among quantum arrangements, mutually orthogonal quantum arrangements, quantum orthogonal arrays and $k$-uniform states.}
\label{fig:a}
\end{figure}

In 2018, Goyeneche et al. \cite{Goyeneche1} generalized some concepts of classical combinatorial designs to quantum combinatorial designs including quantum arrangements: quantum Latin square (QLS), quantum Latin cube (QLC), quantum Latin hypercube (QLH); and mutually orthogonal quantum arrangements: mutually orthogonal quantum Latin squares (MOQLS), mutually orthogonal quantum Latin cubes (MOQLC), mutually orthogonal quantum Latin hypercubes (MOQLH) and quantum orthogonal array (QOA), together with their close relationships and connections to $k$-uniform states (see Fig.\ref{fig:a}). Especially, they derived infinite classes of quantum orthogonal arrays of strength 2 with levels of prime number, thus 2-uniform states with dimension of prime number were obtained. Afterwards, Zang et al. introduced some new constructions on MOQLS and MOQLC, and then 2 and 3-uniform states with finite systems were derived \cite{Zang2}.

In this paper, by following the work of Goyeneche et al. \cite{Goyeneche1}, we will establish infinite families of quantum orthogonal arrays of strength 3 with levels of prime power, and then present the corresponding 3-uniform states. Interestingly, these states we construct are not locally equivalent to any currently known 3-uniform states.

This paper is organized as follows. In Section~\ref{sec:QOA}, we introduce the concepts of orthogonal array and quantum orthogonal array, and their connections to $k$-uniform states. In Section~\ref{sec:qubits}, we present the constructions of quantum orthogonal arrays of strength 3 with 2 levels and $N$ factors for arbitrary $N\geq 6$ and $N\neq 7,8,9,11$. In Section~\ref{qudits}, we depict the detailed constructions of quantum orthogonal arrays of strength 3 with levels of any prime power $d\geq 7$ and $N$ factors for arbitrary $N\geq 7$. In addition, we give quantum orthogonal arrays of strength 3 with 3 and 5 levels and $N$ factors for arbitrary $N\geq 7$, $N\equiv 1$ (mod 3). In Section~\ref{conclusion}, we extend the results of quantum orthogonal arrays of strength 2 with levels of prime number in Ref. \cite{Goyeneche1} to the ones with levels of prime power. Finally we draw our conclusions and outlook.

For simplicity we denote in the following $[d]=\{0,1,\ldots,d-1\}$, $I_d$ the identity matrix of order $d$, $\mathbb{F}_d=\{\alpha_0=0,\alpha_1=1,\alpha_2,\dots, \alpha_{d-1}\}$ the field of order~$d$ with the multiplication group $\mathbb{F}^*_d=\{\alpha_1=1,\alpha_2,\dots, \alpha_{d-1}\}$. For convenience, we also denote the element $\alpha_{i}$ in $\mathbb{F}_d$ by its subscript $i$ for $i\in [d]$.

\section{quantum orthogonal arrays}
\label{sec:QOA}

We first recall the concepts of orthogonal array and quantum orthogonal array, and the connections between the orthogonal array, quantum orthogonal array and the $k$-uniform states.

An orthogonal array of size $r$, with $N$ factors, $d$ levels, and strength $k$, denoted by OA$(r,N,d,k)$, is a $r \times N$ array over a set $V$ of $d$ symbols such that each of its $r\times k$ subarrays contains every $k$-tuple from $V^{k}$ exactly $\lambda=\frac{r}{d^k}$ times. An orthogonal array OA$(r,N,d,k)$ is called irredundant (IrOA), if in any $r\times (N-k )$ subarray, all of its rows are different. By the definitions, one can easily get that an OA$(r,N,d,k)$ is also an OA$(r,N,d,k-1)$ and an OA$(d^k,N,d,k)$ is just an IrOA$(d^k,N,d,k)$.

\begin{definition}
A quantum orthogonal array QOA$(r,N,d,k)$ is an arrangement consisting of $r$ rows composed by $N$-partite  pure quantum states $|\varphi_{i}\rangle\in \mathbb{C}_d^{ N}$
if
\begin{eqnarray}
\sum_{i,j\in[r]}Tr_{l_{1},\ldots,l_{N-k}}(|\varphi_{i}\rangle \langle \varphi_{j}|)
&=\frac{r}{d^{k}} I_{d^{k}}
\end{eqnarray}
for every subset $\{l_{1},\ldots,l_{N-k}\}$ of $N-k$ parties.
\end{definition}

A QOA$(r,N,d,k)$ is obtained when we write each row $(x_1,x_2,\ldots,x_N)$ in an IrOA$(r,N,d,k)$ as a quantum item $|x_1,x_2,\ldots,$ $x_N\rangle$. Moreover, a QOA$(r,N,d,k)$ can give rise to a $N$-partite $k$-uniform  state $|\Phi\rangle$, with local dimension $d$ and $r$ items, i.e., $|\Phi\rangle=\frac{1}{\sqrt{r}}\sum\limits_{i\in[r]}|\varphi_{i}\rangle$, where $|\varphi_{i}\rangle$ is the $i$th row item of the QOA$(r,N,d,k)$. To illustrate these ideas let us consider the following example:
\begin{center}
\footnotesize
 \setlength{\arraycolsep}{5 pt}
 \renewcommand\arraystretch{0.6}
   $
\text{IrOA}(8,6,2,2)=\small{\left(
\begin{array}{cccccc}
0 & 0& 0& 0& 0& 0\\
0 & 0& 1& 1& 1& 0\\
0 & 1& 1& 1& 0& 1 \\
0 & 1& 0& 0& 1& 1\\
1 & 0& 1& 0& 1& 1\\
1 & 0& 0& 1& 0& 1\\
1 & 1& 0& 1& 1& 0\\
1 & 1& 1& 0& 0& 0
\end{array}
\right),}$
\end{center}
from which we get directly a QOA,
\begin{center}
\footnotesize
 \setlength{\arraycolsep}{6 pt}
 \renewcommand\arraystretch{0.6}
   $
\text{QOA}(8,6,2,2)=\small{\left(
\begin{array}{c}
|0~0~0~0~0~0\rangle\\
|0~0~1~1~1~0\rangle\\
|0 ~1~ 1~ 1~ 0~ 1\rangle \\
|0~  1~ 0~ 0~ 1~ 1\rangle\\
|1 ~ 0~ 1~ 0~ 1~ 1\rangle\\
|1  ~0~ 0~ 1~ 0~ 1\rangle\\
|1  ~1~ 0~ 1~ 1~ 0\rangle\\
|1  ~1~ 1~ 0~ 0~ 0\rangle
\end{array}
\right).}$
\end{center}
The above can be written as for convenience,
\begin{center}
\footnotesize
 \setlength{\arraycolsep}{5 pt}
 \renewcommand\arraystretch{0.8}
   $\text{QOA}(4,6,2,2)=\small{\left(
\begin{array}{ccc}
|0\rangle & |0\rangle& (|0000\rangle+|1110\rangle)/\sqrt{2}\\
|0\rangle & |1\rangle&(|1101\rangle+|0011\rangle)/\sqrt{2} \\
|1\rangle & |0\rangle& (|1011\rangle+|0101\rangle)/\sqrt{2}\\
|1\rangle & |1\rangle&(|0110\rangle+|1000\rangle)/\sqrt{2}
\end{array}
\right).}$
\end{center}
The corresponding 2-uniform state for 6-qubit systems is then given by
\begin{eqnarray}
\begin{array}{cl}
|\Phi\rangle =&\frac{1}{2\sqrt{2}}(|000000\rangle +|001110\rangle+|011101\rangle+|010011\rangle\\
 &~~~+|101011\rangle +|100101\rangle+|110110\rangle+|111000\rangle).
\end{array}
\end{eqnarray}

\section{Quantum orthogonal arrays of strength 3 with 2 levels}
\label{sec:qubits}

We now focus on the constructions of QOAs of strength 3 with 2 levels. We apply the `Cl $+$ Q' method mentioned in Ref. \cite{Raissi}. In detail, we utilize classical orthogonal arrays instead of the part of `Cl'; GHZ states and Bell states instead of the part of `Q' here.

It is well known that 3-qubit GHZ state is 1-uniform but also an AME(3,2) state. Goyeneche et al. introduced a class of GHZ states that can form an orthonormal basis of ~$\mathbb{C}_2 ^3$ in \cite{Goyeneche1}. Here we examine their further properties and list the results in Appendix \ref{sec:AppF}.
\begin{lemma} \ {\rm (\cite{Goyeneche1})}\label{ghz} \ Let
$|\text{GHZ}_{ijk}\rangle=(-1)^{\alpha_{ijk}}\sigma_{i}
\otimes\sigma_{j}\otimes\sigma_{k}|\text{GHZ}\rangle$, where
$\alpha_{ijk}=1$ if $i=j=k$ and $0$ otherwise, $\sigma_{0}$, $\sigma_{1}$ represent the Pauli matrices~$\sigma_{x}$, $\sigma_{z}$, respectively. Then $|\text{GHZ}_{ijk}\rangle$ is a $1$-uniform state and $\{|\text{GHZ}_{ijk}\rangle:i,j,k\in \{0,1\}\}$ constitutes an orthonormal basis of ~$\mathbb{C}_2 ^3$.
\end{lemma}

For any $i,j,k\in \{0,1\}$, $|\text{GHZ}_{ijk}\rangle$ can actually be written as
\begin{eqnarray}|\text{GHZ}_{ijk}\rangle&=(-1)^{\alpha_{ijk}}
\frac{1}{\sqrt{2}}\sigma_{i}\otimes\sigma_{j}\otimes\sigma_{k}(|000\rangle+|111\rangle)\nonumber \\
&\hspace{-0.3cm}=(-1)^{\alpha_{ijk}}\frac{1}{\sqrt{2}}
\{|\bar{i}\bar{j}\bar{k}\rangle+(-1)^{\omega(ijk)}|ijk\rangle\},
\end{eqnarray}
where $\bar{s}=1-s$, $s\in \{i,j,k\}$ and $\omega(ijk)$ is the number of 1s in the multiset $\{i,j,k\}$.
\begin{construction}\mbox{}\hspace{-8pt}\label{con1} \ {\rm (\cite{Seiden})}
For any given $k$, all possible $2^{k+1}$ $(k+1)$-tuple columns can form an OA$(2^{k+1},k+1,2,k)$.
\end{construction}

Based on the above conclusions, we have the main results.
\begin{theorem}\label{02}
For any integer $m\geq 1$ and $m\neq 2$, there exists a QOA$(8,3+3m,2,3)$.
\end{theorem}
\noindent \p Define $|\Phi_{ijk}\rangle\in \mathbb{C}_2^{3+3m}$, $|\Phi_{ijk}\rangle=|ijk\rangle|\text{GHZ}_{ijk}\rangle^{\otimes m}, $ where $i,j,k\in \{0,1\}$.
Then  $\{|\Phi_{ijk}\rangle:i,j,k\in \{0,1\}\}$ forms a QOA$(8,3+3m,2,3)$  for any $m\geq 1$ and $m\neq 2$. The detailed proof is given in Appendix \ref{sec:AppA}.
\qed

In Theorem \ref{02} the case $m=2$ is excluded. This fact can be seen, for instance, by taking $S\cap C=\{1\}$, $S\cap Q_1=\{4\}$ and $S\cap Q_2=\{7\}$. We have
\begin{eqnarray*}
&\hspace{-4.4cm}\rho_{S}
=\sum\limits_{i,i',j,k=0}^{1}|i\rangle\langle i'| \otimes \{Tr_{2,3} |\text{GHZ}_{ijk}\rangle \langle \text{GHZ}_{i'jk}|\}^{\otimes 2}\\
&=\frac{1}{4}\sum\limits_{i,i',j,k=0}^{1}|i\rangle\langle i'| \otimes\{ (-1)^{\alpha_{ijk}+\alpha_{i'jk}} [|\bar{i}\rangle\langle\bar{i'}|
+(-1)^{\omega(i'jk)+\omega(ijk)}|i\rangle\langle i'|]\}^{\otimes 2}\\
&\hspace{-2.4cm}=\frac{1}{4}\sum\limits_{i,i',j,k=0}^{1}|i\rangle\langle i'| \otimes\{|\bar{i}\rangle\langle\bar{i'}|
+(-1)^{\omega(i'jk)+\omega(ijk)}|i\rangle\langle i'|\}^{\otimes 2}.
\end{eqnarray*}
\noindent If~$i\neq i'$, then $\rho_{S}=\frac{1}{4}\sum\limits_{i,j,k=0}^{1}|i\rangle\langle \bar{i}| \otimes\{ |\bar{i}\rangle\langle i|
-|i\rangle\langle \bar{i}|\}^{\otimes 2}\neq 0$. Hence, ~$\rho_{S}\neq \mathbb{I}_8$.

\begin{construction}\mbox{}\hspace{-8pt}\label{con2} \ {\rm (\cite{Hedayat})}
For any $l\in [d]$, the array consisting of $(x_1,x_2,\ldots, x_N)$ as the row satisfying $x_1+x_2+\cdots+x_N\equiv l$ \rm{(mod} $d)$    can constitute an OA$(d^{N-1},N,d,N-1)$, where $x_1,x_2,\ldots,x_N\in[d]$.
\end{construction}

From the Construction \ref{con2}, it is easy to find that $(i,j,k,i+j+k)$ as a row can form an OA$(8,4,2,3)$ for $i,j,k\in\{0,1\}$. Let us add one item $i+j+k$ to the state $|\Phi_{ijk}\rangle$ in Theorem \ref{02}. As a result we obtain a QOA$(8,4+3m,2,3)$. However, since AME(7,2) does not exist \cite{Huber1}, it is no surprising that $m\neq 1$.

\begin{theorem}\label{002}
For any integer $m\geq 2$, there exists a QOA$(8,4+3m,2,3)$.
\end{theorem}

\noindent \p Define $|\Phi_{ijk}\rangle\in \mathbb{C}_2^{4+3m}$,
$|\Phi_{ijk}\rangle=|i,j,k,i+j+k\rangle|\text{GHZ}_{ijk}\rangle^{\otimes m}$, where $i,j,k\in \{0,1\}, m\geq 2$. It is easy to check that $\{|\Phi_{ijk}\rangle:i,j,k\in \{0,1\}\}$ forms a QOA$(8,4+3m,2,3)$. Since $m\geq 2$ and $(i,j,k,i+j+k)$ as a row can form an OA$(8,4,2,3)$ for $i,j,k\in\{0,1\}$, the Eq. (\ref{01}) in Appendix \ref{sec:AppA} holds for any $|S|=3$.

According to Lemma \ref{ghz} and Lemma \ref{ghz2}, $\rho_{S}= \mathbb{I}_8$ is true except for the cases $S\cap C=\{4\}$ and $|S\cap Q_1|=2$. Without of generality suppose $S\cap Q_1=\{5,6\}$. We have
\begin{eqnarray*}
&\hspace{-0.4cm}\rho_{S}
=\sum\limits_{i,j,k=0}^{1}|i+j+k\rangle\langle i+j+k| \otimes Tr_{3} |\text{GHZ}_{ijk}\rangle \langle \text{GHZ}_{ijk}|\\
&=\frac{1}{2}\sum\limits_{i,j,k=0}^{1}|i+j+k\rangle\langle i+j+k| \otimes\{|\overline{ij}\rangle\langle\overline{ij}|
+|ij\rangle\langle ij|\}\\
&\hspace{-3.7cm}=\frac{1}{2}I_2 \otimes \sum\limits_{i,j=0}^{1}\{|\overline{ij}\rangle\langle\overline{ij}|
+|ij\rangle\langle ij|\}\\
&\hspace{-8.3cm}=I_8,
\end{eqnarray*}

\noindent which completes the proof.
\qed
For the Bell states $|\Phi^{\pm}\rangle=\frac{1}{\sqrt{2}}(|00\rangle\pm |11\rangle)$ and $|\Psi^{\pm}\rangle=\frac{1}{\sqrt{2}}(|01\rangle\pm |10\rangle)$,  we can  rewrite them as
\begin{eqnarray}
|\phi_{xy}\rangle=\frac{1}{\sqrt{2}}\{(-1)^x|xy\rangle+|\bar{x}\bar{y}\rangle\},~~ x,y\in\{0,1\}.
\end{eqnarray}
The properties of Bell states in the new form are presented in Appendix \ref{sec:AppG}.

Combining the Construction \ref{con1} with Lemmas \ref{bell1}-\ref{bell3}, we get the last main result of this section.
\begin{theorem}\label{03}
For any integer $m\geq 1$, there exists a QOA$(32,11+3m,2,3)$.
\end{theorem}

\noindent \p We can prove the theorem by consider $|\Phi_{ijkfg}\rangle\in \mathbb{C}_2^{11+3m}$,
\begin{eqnarray}
|\Phi_{ijkfg}\rangle=|ijkfg\rangle|\phi_{f+i,g+j}\rangle|\phi_{g+k,f+i}
\rangle|\phi_{g+j,i+k}\rangle|\text{GHZ}_{jfg}\rangle^{\otimes m},
\end{eqnarray}
where $i,j,k,f,g\in \{0,1\}, m\geq 1$. The proof that $\{|\Phi_{ijkfg}\rangle:i,j,k,f,g\in \{0,1\}\}$ forms a QOA$(32,11+3m,2,3)$ is shown in Appendix \ref{sec:AppB}.
\qed

\section{Quantum orthogonal arrays of strength 3 with levels of prime power}
\label{qudits}
We now devote to study the constructions of QOAs of strength 3 with levels of prime power. We adopt the `Cl $+$ Q' method mentioned in Ref. \cite{Raissi} with some results of classical orthogonal arrays. Let us start from an orthonormal basis of~$\mathbb{C}_d^{3}$ for any prime power $d\geq 3$.

\begin{lemma}\label{phi} Let
\begin{equation}
|\psi_{ijk}\rangle=\frac{1}{\sqrt{d}}\sum\limits_{l\in[d]}\omega^{il}|l+j,l+\alpha j+\beta k,l\rangle,
\end{equation}
where $d\geq 3$ is a prime power, $\omega=e^{\frac{2\pi \sqrt{-1}}{d}}$, $\alpha,\beta\in \mathbb{F}^*_d$, $\alpha\neq 1$ and the operations in kets
are taken in $\mathbb{F}_d$. Then $|\psi_{ijk}\rangle$ is a $1$-uniform state for arbitrary $i,j,k\in [d]$. Moreover, $\{|\psi_{ijk}\rangle:i,j,k\in[d]\}$ constitutes an orthonormal basis of~$\mathbb{C}_d^{3}$.
\end{lemma}

\noindent \p Firstly, $|\psi_{ijk}\rangle$ is 1-uniform for any $i,j,k\in [d]$ since
\begin{eqnarray*}
&\hspace{-9cm}\rho_1=Tr_{23}|\psi_{ijk}\rangle\langle\psi_{ijk}|\\
&=\frac{1}{d}\sum\limits_{l,l'\in[d]}Tr_{23}\{\omega^{i(l-l')}|l+j,l+\alpha  j+\beta k,l\rangle \langle l'+j,l'+\alpha j+\beta k,l'|\}\\
&\hspace{-8.1cm}=\frac{1}{d}\sum\limits_{l\in[d]}|l+j\rangle \langle l+j|\\
&\hspace{-10.8cm}=\frac{I_d}{d}.
\end{eqnarray*}
\noindent Similar results hold for the other cases. Secondly, we show the orthogonality as follows:
\begin{eqnarray*}
&\langle\psi_{i'j'k'}|\psi_{ijk}\rangle=\frac{1}{d}\sum\limits_{l,l'\in [d]} \omega^{il-i'l'}\langle l'+j',l'+\alpha j'+\beta k',l'|l+j,l+\alpha j+\beta k,l\rangle\\
&\hspace{1cm}=\frac{1}{d}\sum\limits_{l\in [d]} \omega^{(i-i')l}\langle l+j',l+\alpha  j'+\beta k'|l+j,l+\alpha  j+\beta k\rangle\\
&\hspace{-6.2cm}=\delta_{i'i}\delta_{j'j}\delta_{k'k},
\end{eqnarray*}
\noindent which ends the proof.
\qed

\begin{lemma}\label{phi1} Let $|\psi_{ijk}\rangle$ be defined in Lemma \ref{phi} for
$i,j,k\in[d]$. Then we have the following equalities:
\begin{enumerate}
\item $\sum\limits_{j\in [d]}Tr_1|\psi_{ijk}\rangle\langle\psi_{ijk}|=\sum\limits_{k\in [d]}Tr_1|\psi_{ijk}\rangle\langle\psi_{ijk}|=\frac{I_{d^2}}{d};$\label{21}
\item $\sum\limits_{j\in [d]}Tr_2|\psi_{ijk}\rangle\langle\psi_{ijk}|=\frac{I_{d^2}}{d};$\label{22}
    \item $\sum\limits_{j\in [d]}Tr_3|\psi_{ijk}\rangle\langle\psi_{ijk}|=\sum\limits_{k\in [d]}Tr_3|\psi_{ijk}\rangle\langle\psi_{ijk}|=\frac{I_{d^2}}{d}$.\label{23}
\end{enumerate}
\end{lemma}

\noindent \p With respect to the item \ref{21}, we have
\begin{eqnarray*}
&\hspace{-4cm}\sum\limits_{j\in [d]}Tr_1|\psi_{ijk}\rangle\langle\psi_{ijk}|\\
&=\frac{1}{d}\sum\limits_{j,l\in [d]}|l+\alpha j+\beta k,l\rangle\langle l+\alpha  j+\beta k,l|=\frac{I_{d^2}}{d}.
\end{eqnarray*}
Note that for any $(x,y)$, the following set of equations has only one solution for any $k\in[d]$,
\begin{equation}
    \left\{
          \begin{array}{ll}
l+\alpha  j+\beta k=x,\\
l=y.\\
          \end{array}
       \right.
\end{equation}

\noindent In fact, $j=\alpha ^{-1}(x-y-\beta k)$, $l=y$ is just the solution to the above equations, where $\alpha^{-1}$ is the inverse element of ~$\alpha$ in group~$\mathbb{F}^*_d$.

Concerning the items \ref{22} and \ref{23}, we have
\begin{eqnarray*}
&\hspace{-1.2cm}\sum\limits_{j\in [d]}Tr_2|\psi_{ijk}\rangle\langle\psi_{ijk}|\\
&=\frac{1}{d}\sum\limits_{j,l\in [d]}|l+j,l\rangle\langle l+j,l|=\frac{I_{d^2}}{d}
\end{eqnarray*}
\noindent and
\begin{eqnarray*}
&\hspace{-5.2cm}\sum\limits_{j\in [d]}Tr_3|\psi_{ijk}\rangle\langle\psi_{ijk}|\\
&=\frac{1}{d}\sum\limits_{j,l\in [d]}|l+j,l+\alpha  j+\beta k\rangle\langle l+j,l+\alpha  j+\beta k|=\frac{I_{d^2}}{d}.
\end{eqnarray*}
Since for any  $(x,y)$, the following set of equations has one solution for any $k\in[d]$,
\begin{equation}
    \left\{
          \begin{array}{ll}
l+j=x,\\
l+\alpha  j+\beta k=y,\\
          \end{array}
       \right.
\end{equation}
which is given by $j=(\alpha  -1)^{-1}(y-\beta k-x)$, $l=[1+(\alpha  -1)^{-1}]x-(\alpha  -1)^{-1}(y-\beta k)$. Similar proofs hold for the rest cases.
\qed

\begin{construction}\mbox{}\hspace{-8pt}\label{04} \ \ {\rm(\cite{Arkin})}
Suppose~$d>3$ is a prime power. Let
\begin{eqnarray}
A_{ijk}=(i,k,i+\alpha_1 j+\alpha_1^2 k,i+\alpha_2 j+\alpha_2^2 k, \dots, i+\alpha_{d-1} j+\alpha_{d-1}^2 k),\label{shizioa}
\end{eqnarray}
where~$i,j,k\in [d]$, $\alpha_l \in \mathbb{F}_d^* ~(l=1,2,\dots,d-1)$ and  the operations are taken in $\mathbb{F}_d$. Then~$A_{ijk}$ as a row constitutes an OA$(d^3,d+1,d,3)$ for $i,j,k\in [d]$; moreover if $d=2^{t}$, then the array after adding $j$ as a column is an OA$(d^3,d+2,d,3)$ for $t\geq 2$.
\end{construction}

Combining the Construction \ref{04} with Lemmas \ref{phi}-\ref{phi1}, we have the following constructions of QOA of strength 3 with levels of prime power. See the proof in Appendix \ref{sec:AppD}.

\begin{construction}\label{qud}
Let $|\Phi_{ijk}\rangle\in \mathbb{C}_d^{4+3m}$, $i,j,k\in [d], m\geq 1$, be defined as follows,
\begin{eqnarray}
|\Phi_{ijk}\rangle=|i,k,i+j+k,i+\alpha_2  j+\alpha_2 ^2k\rangle|\psi_{ijk}\rangle^{\otimes m},\label{40}
\end{eqnarray}
where $\alpha_2 \in \mathbb{F}^*_d$, $\alpha_2 ^2\neq 1$, $|\psi_{ijk}\rangle$ is defined in Lemma \ref{phi} with the elements $\alpha=\alpha_2$, $\beta \neq \alpha_2, \alpha_2 ^2$, $ \alpha_2 -1$,  $\alpha_2 ^2-1$ and $\alpha_2 ^2\pm\alpha_2$, and the operations in kets are taken in $\mathbb{F}_d$. Then $\{|\Phi_{ijk}\rangle:i,j,k\in [d]\}$ forms a QOA$(d^3,4+3m,d,3)$.
\end{construction}


Let us add some items $i+\alpha_{3} j+\alpha_{3}^2 k, i+\alpha_{4} j+\alpha_{4}^2 k,\dots, i+\alpha_{n} j+\alpha_{n}^2 k$ to Eq. (\ref{40}), where~$\alpha_{3}, \alpha_{4}, \dots, \alpha_{n}$ are all different and not equal to $\alpha_{1}=1$ and $\alpha_2$. There is no doubt that $\beta\neq \alpha_{s}\alpha_{2}, \alpha_{s}(\alpha_2-1)$ for $s\in \{3,4,\ldots,n\}$, as is shown in the proof of Case 3 in Construction \ref{qud}. In consequence, we have a new QOA.

\begin{construction}\label{qud2}
Let $|\Phi_{ijk}\rangle\in \mathbb{C}_d^{2+n+3m}$, $i,j,k\in [d], n\geq 2$, $m\geq 1$, be defined as follows,
\begin{eqnarray}
|\Phi_{ijk}\rangle=|i,k,i+\alpha_{1}j+\alpha^2_{1}k,\dots,i+\alpha_s j+\alpha_s^2 k, \dots, i+\alpha_{n} j+\alpha_{n}^2 k\rangle |\psi_{ijk}\rangle^{\otimes m},
\end{eqnarray}
where $\alpha_2 \in \mathbb{F}^*_d$, $\alpha_2 ^2\neq 1$,  $|\psi_{ijk}\rangle$ is defined in Lemma \ref{phi} with the elements $\alpha=\alpha_2$, $\beta \neq  \alpha_2^2-1$, $\alpha_2^2+\alpha_2$, $\alpha_{2}\alpha_{s}$ and $(\alpha_{2}-1)\alpha_{s}$ for $s\in\{1,2,\ldots,n\}$, and the operations in kets
are taken in $\mathbb{F}_d$. Then $\{|\Phi_{ijk}\rangle:i,j,k\in [d]\}$ forms a QOA$(d^3,2+n+3m,d,3)$.
\end{construction}

\begin{theorem}\label{qoa3}
For any prime power $d\geq 7$ and integer $N\geq 7$, there exists a QOA$(d^3,N,d,3)$.
\end{theorem}

\noindent \p Clearly, for $d\geq 16$ there must exist $\beta $ satisfying the condition for $n=2,3,4$ in Construction \ref{qud2}. Thus the conclusion establishes for any $d\geq 16$. Furthermore, when $d=7,11,13$, set $\alpha_2=2$, $\alpha_3=3$, $\alpha_4=4$ and $\beta =5$ in Construction \ref{qud2}. As a result, the conclusion holds. When $d=8$, let $\alpha_2=x$, $\alpha_3=x^2$, $\alpha_4=x+1$ and $\beta =x^2+x+1$ in the field $\mathbb{F}_8=\{0,1,x,x+1,x^2,x^2+1,x^2+x,x^2+x+1\}$ with irreducible polynomial $x^3+x^2+1$.  When $d=9$, assume $\alpha_2=x+1$, $\alpha_3=x$, $\alpha_4=2x$ and $\beta =x+2$ in the field $\mathbb{F}_9=\{0,1,2,x,x+1,x+2,2x,2x+1,2x+2\}$ with irreducible polynomial $x^2+2$ \cite{Lidl}. The conclusion can be obtained from Construction \ref{qud2}.
\qed

According to Construction \ref{con2}, it is not difficult to verify that $(i,j,k,(d-1)(i+j+k))$ as a row forms an OA$(d^{3},4,d,3)$ for $i,j,k\in[d]$. As a consequence, we get the following result.

\begin{theorem}\label{qud3}
For any $m\geq 1$, there exists a QOA$(27,4+3m,3,3)$ and a QOA$(125,4+3m,5,3)$.
\end{theorem}

\noindent \p Let $|\Phi_{ijk}\rangle\in \mathbb{C}_d^{4+3m}$, $i,j,k\in [d]$, $d\in\{3,5\}$, $m\geq 1$, be defined as follows,
\begin{eqnarray}
|\Phi_{ijk}\rangle=|i,j,k,(d-1)(i+j+k)\rangle|\phi_{ijk}\rangle^{\otimes m},
\end{eqnarray}
where
\begin{eqnarray}
|\phi_{ijk}\rangle=\frac{1}{\sqrt{d}}\sum\limits_{l\in[d]}\omega^{il}|l+2j+k,l+j+2k,l\rangle,
\end{eqnarray}
and the sums in kets are taken modulo $d$. Then it is not difficult to check that $\{|\Phi_{ijk}\rangle:i,j,k\in [d]\}$ forms a QOA$(d^3,4+3m,d,3)$ for $d=3,5$.
\qed

\section{Conclusions}
\label{conclusion}

As listed in Lemma \ref{oa0} in Appendix \ref{sec:AppE} the authors in \cite{Goyeneche1} presented some constructions for a series of infinite classes of quantum orthogonal arrays of strength 2 with levels of any prime number. These constructions for odd prime number $d$ can be extended to that for any prime power $d\geq 3$. It is straightforward to prove the following conclusion.

\begin{lemma}\label{0a333}
For any prime power $d\geq 3$ and $m\geq1$, define
$|\varphi^1_{ij}\rangle=|i,j,i+j \rangle |\phi_{i,j}\rangle^{\otimes m}$ and
$|\varphi^2_{ij}\rangle=|i,j,i+j,i+\alpha j \rangle |\phi_{i,j}\rangle^{\otimes m}$, where $\alpha\neq1\in \mathbb{F}^*_d$, $|\phi_{i,j}\rangle=\frac{1}{\sqrt{d}}\sum\limits_{l\in[d]}\omega^{il}|l+j,l\rangle$ with $\omega=e^{\frac{2\pi\sqrt{-1}}{d}}$ and the operations in kets
are taken in $\mathbb{F}_d$. Then $\{|\varphi^1_{ij}\rangle:i,j\in [d]\}$ forms a QOA$(d^2,3+2m,d,2)$ and $\{|\varphi^2_{ij}\rangle:i,j\in [d]\}$ forms a QOA$(d^2,4+2m,d,2)$.
\end{lemma}

In conclusion, we have generalize the constructions of a series of infinite classes of quantum orthogonal arrays of strength 2 with levels of prime number \cite{Goyeneche1} to the ones with levels of prime power on the one hand. On the other hand, we have explicitly presented the construction methods for some infinite classes of quantum orthogonal arrays of strength 3 with levels of prime power.
On account of the connection between a quantum orthogonal array and a $k$-uniform state as we mentioned in Section \ref{sec:QOA}, combining Lemma \ref{oa0} and Lemma \ref{0a333} one has

\begin{theorem}
For any prime power $d\geq 2$ and $N\geq 5$, there exists a QOA$(d^2,N,d,2)$. Correspondingly a 2-uniform state exists for $N$ systems with dimension $d$.
\end{theorem}

In addition, the Theorems \ref{02}, \ref{002}-\ref{03} and Theorem  \ref{qoa3} can be integrated as follows.

\begin{theorem}
For any $N\geq 6$ and $N\neq 7,8,9,11$, there exists a QOA$(r,N,2,3)$, where $r=8$ when $N\equiv 0,1$ {\rm (mod 3)} and $r=32$ when $N\equiv 2$ {\rm (mod 3)}. Correspondingly a 3-uniform state exists for $N$-qubit systems.
\end{theorem}
\begin{theorem}
For any prime power $d\geq 7$ and $N\geq 7$, there exists a QOA$(d^3,N,d,3)$; and a 3-uniform state correspondingly exists for $N$ systems with dimension $d$.
\end{theorem}

Although there is some work on higher uniformity $k\geq 4$ of $k$-uniform states \cite{Pang2,Chen}, however it is far from enough. On the other side,  there is less work on orthogonal arrays with strength $k\geq 4$ \cite{Stinson,Bierbrauer,Mukhopadhyay}. It would be of great interest to consider the constructions of $k$-uniform states via quantum orthogonal arrays with higher strength $k\geq 4$.

\section*{Acknowledgements}
This work is supported by Beijing Postdoctoral Research Foundation (2022ZZ071), Natural Science Foundation of Hebei Province (F2021205001), NSFC (Grant Nos. 11871019, 12075159, 12171044, 62272208), Beijing Natural Science Foundation (Z190005), Academician Innovation Platform of Hainan Province.

\appendix
\section{The properties of GHZ states}
\label{sec:AppF}

In the  Lemmas \ref{ghz1}-\ref{ghz4}, we always assume that $i,j,k,i',j',k' \in \{0,1\}$.
\begin{lemma}\label{ghz1}
\begin{enumerate}
\item $\sum\limits^{1}_{j=0}Tr_{23}|\text{GHZ}_{ijk}\rangle\langle\text{GHZ}_{i'jk}|=\sum\limits^{1}_{k=0}Tr_{23}|\text{GHZ}_{ijk}\rangle\langle\text{GHZ}_{i'jk}|=I_2\delta_{ii'};$
\item $\sum\limits^{1}_{i=0}Tr_{13}|\text{GHZ}_{ijk}\rangle\langle\text{GHZ}_{ij'k}|=\sum\limits^{1}_{k=0}Tr_{13}|\text{GHZ}_{ijk}\rangle\langle\text{GHZ}_{ij'k}|=I_2\delta_{jj'};$
\item
    $\sum\limits^{1}_{i=0}Tr_{12}|\text{GHZ}_{ijk}\rangle\langle\text{GHZ}_{ijk'}|=\sum\limits^{1}_{j=0}Tr_{12}|\text{GHZ}_{ijk}\rangle\langle\text{GHZ}_{ijk'}|=I_2\delta_{kk'}$.
\end{enumerate}
\end{lemma}
\noindent \p For the item \ref{000}, we have
\begin{eqnarray*}
&\hspace{-5.8cm}\sum\limits^{1}_{j=0}Tr_{23}|\text{GHZ}_{ijk}\rangle\langle\text{GHZ}_{i'jk}|\\
&\hspace{-2cm}=\frac{1}{2}\sum\limits^{1}_{j=0}(-1)^{\alpha_{ijk}+\alpha_{i'jk}}\{|\bar{i}\rangle\langle\bar{i'}|+(-1)^{\omega(ijk)+\omega(i'jk)}|i\rangle\langle i'|\}\\
&=\frac{1}{2}[(-1)^{\alpha_{ikk}+\alpha_{i'kk}}+(-1)^{\alpha_{i\bar{k}k}+\alpha_{i'\bar{k}k}}]\{|\bar{i}\rangle\langle\bar{i'}|+(-1)^{\omega(i)+\omega(i')}|i\rangle\langle i'|\}\\
&\hspace{-9.7cm}=I_2\delta_{ii'},
\end{eqnarray*}
where the last equality holds since if $i\neq i'$, then $(-1)^{\alpha_{ikk}+\alpha_{i'kk}}+(-1)^{\alpha_{i\bar{k}k}+\alpha_{i'\bar{k}k}}=0$ for any $k\in\{0,1\}$.
Similar proofs hold for the rest cases.\qed

\begin{lemma}\label{ghz3}
\begin{enumerate}
\item $\sum\limits^{1}_{i=0}Tr_{23}|\text{GHZ}_{ijk}\rangle\langle\text{GHZ}_{ij'k'}|=I_2\delta_{jj'}\delta_{kk'};$
\item $\sum\limits^{1}_{j=0}Tr_{13}|\text{GHZ}_{ijk}\rangle\langle\text{GHZ}_{i'jk'}|=I_2\delta_{ii'}\delta_{kk'};$
\item $\sum\limits^{1}_{k=0}Tr_{12}|\text{GHZ}_{ijk}\rangle\langle\text{GHZ}_{i'j'k}|=I_2\delta_{ii'}\delta_{jj'}$.
\end{enumerate}
\end{lemma}
\noindent \p We prove the first equality of item \ref{000}.
\begin{eqnarray*}
&\hspace{-6cm}\sum\limits^{1}_{i=0}Tr_{23}|\text{GHZ}_{ijk}\rangle\langle\text{GHZ}_{ij'k'}|\\
&=\frac{1}{2}\sum\limits^{1}_{i=0}(-1)^{\alpha_{ijk}+\alpha_{ij'k'}} Tr_{23} [|\bar{i}\bar{j}\bar{k}\rangle\langle\bar{i}\bar{j'}\bar{k'}|
+(-1)^{\omega(ij'k')+\omega(ijk)}|ijk\rangle\langle ij'k'|
\\
&\hspace{-2.4cm}+(-1)^{\omega(ijk)}|ijk\rangle\langle\bar{i}
\bar{j'}\bar{k'}|+(-1)^{\omega(ij'k')}|\bar{i}\bar{j}\bar{k}\rangle\langle ij'k'|]
\\
&\hspace{-10.3cm}=I_2\delta_{ii'},
\end{eqnarray*}
where the last equality is due to that if $j\neq j' $and $k\neq k'$, then
\begin{eqnarray*}
\sum\limits_{i=0}^{1}(-1)^{\alpha_{ijk}+\alpha_{i\bar{j}\bar{k}}} \{(-1)^{\omega(ijk)}|i\rangle\langle \bar{i}|+(-1)^{\omega(i\bar{j}\bar{k})}|\bar{i}\rangle\langle i|\}=0
\end{eqnarray*}
for any $j,k\in\{0,1\}$.
Similar proofs apply to the rest cases.\qed

\begin{lemma}\label{ghz2}
\begin{enumerate}
\item $\sum\limits^{1}_{j=0}Tr_{1}|\text{GHZ}_{ijk}\rangle\langle\text{GHZ}_{ijk}|
    =\sum\limits^{1}_{k=0}Tr_{1}|\text{GHZ}_{ijk}\rangle\langle\text{GHZ}_{ijk}|
    =\frac{1}{2}I_4;$\label{000}
\item $\sum\limits^{1}_{i=0}Tr_{2}|\text{GHZ}_{ijk}\rangle\langle\text{GHZ}_{ijk}|
    =\sum\limits^{1}_{k=0}Tr_{2}|\text{GHZ}_{ijk}\rangle\langle\text{GHZ}_{ijk}|
    =\frac{1}{2}I_4;$
\item $\sum\limits^{1}_{i=0}Tr_{3}|\text{GHZ}_{ijk}\rangle\langle\text{GHZ}_{ijk}|
    =\sum\limits^{1}_{j=0}Tr_{3}|\text{GHZ}_{ijk}\rangle\langle\text{GHZ}_{ijk}|
    =\frac{1}{2}I_4$.
\end{enumerate}
\end{lemma}
\noindent \p Here, we just prove the item \ref{000}. The other cases are similarly proved.
For any $k\in\{0,1\}$, we have
\begin{eqnarray*}
&\hspace{-3.1cm}\sum\limits^{1}_{j=0}Tr_{1}|\text{GHZ}_{ijk}\rangle\langle\text{GHZ}_{ijk}|\\
&\hspace{-3.45cm}=\frac{1}{2}\sum\limits^{1}_{j=0}|\bar{j}\bar{k}\rangle\langle\bar{j}\bar{k}|+|jk\rangle\langle jk|\\
&=\frac{1}{2}|\bar{k}\bar{k}\rangle\langle\bar{k}\bar{k}|+|kk\rangle\langle kk|+|k\bar{k}\rangle\langle k\bar{k}|+|\bar{k}k\rangle\langle \bar{k}k|\\
&\hspace{-1.2cm}=\frac{1}{2}(|\bar{k}\rangle\langle\bar{k}|+|k\rangle\langle k|)\otimes(|\bar{k}\rangle\langle\bar{k}|+|k\rangle\langle k|)\\
&\hspace{-6.9cm}=\frac{1}{2}I_4.
\end{eqnarray*} \qed
\begin{lemma}\label{ghz5}
\begin{enumerate}
\item
$\sum\limits^{1}_{i,j=0}Tr_{1}|\text{GHZ}_{ijk}\rangle\langle\text{GHZ}_{ijk'}|=\sum\limits^{1}_{i,j=0}Tr_{2}|\text{GHZ}_{ijk}\rangle\langle\text{GHZ}_{ijk'}|=I_4\delta_{kk'};$
\item
$\sum\limits^{1}_{i,k=0}Tr_{1}|\text{GHZ}_{ijk}\rangle\langle\text{GHZ}_{ij'k}|=\sum\limits^{1}_{i,k=0}Tr_{3}|\text{GHZ}_{ijk}\rangle\langle\text{GHZ}_{ij'k}|=I_4\delta_{jj'};$
 \item
$\sum\limits^{1}_{j,k=0}Tr_{2}|\text{GHZ}_{ijk}\rangle\langle\text{GHZ}_{i'jk}|=\sum\limits^{1}_{j,k=0}Tr_{3}|\text{GHZ}_{ijk}\rangle\langle\text{GHZ}_{i'jk}|=I_4\delta_{ii'}$.
\end{enumerate}
\end{lemma}

\noindent \p We prove the item \ref{000} in detail. The other cases can be proved  in the same way. When $k=k'$, the conclusion is true in the light of Lemma \ref{ghz2}. When $k\neq k'$, we have
\begin{eqnarray*}
&\hspace{-4.8cm}\sum\limits^{1}_{i,j=0}Tr_{1}|\text{GHZ}_{ijk}\rangle\langle\text{GHZ}_{ij\bar{k}}|\\
&=\frac{1}{2}\sum\limits_{i,j=0}^{1} (-1)^{\alpha_{ijk}+\alpha_{ij\bar{k}}}  [
|\bar{j}\bar{k}\rangle\langle\bar{j}k|
+(-1)^{\omega(ijk)+\omega(ij\bar{k})}|jk\rangle\langle j\bar{k}|]\\
&\hspace{-2cm}=\frac{1}{2}\sum\limits_{j=0}^{1}\{\sum\limits_{i=0}^{1} (-1)^{\alpha_{ijk}+\alpha_{ij\bar{k}}} \} [
|\bar{j}\bar{k}\rangle\langle\bar{j}k|
-|jk\rangle\langle j\bar{k}|]\\
&\hspace{-9.1cm}=0,
\end{eqnarray*}
\noindent as $\sum\limits_{i=0}^{1} (-1)^{\alpha_{ijk}+\alpha_{ij\bar{k}}}=0$ for any $j\in \{0,1\}$.
\qed

\begin{lemma}\label{ghz4}
\begin{enumerate}
\item
 $\sum\limits^{1}_{j,k=0}Tr_{1}|\text{GHZ}_{ijk}\rangle\langle\text{GHZ}_{i'jk}|=I_4\delta_{ii'};$
 \item
 $\sum\limits^{1}_{i,k=0}Tr_{2}|\text{GHZ}_{ijk}\rangle\langle\text{GHZ}_{ij'k}|=I_4\delta_{jj'};$
 \item
$\sum\limits^{1}_{i,j=0}Tr_{3}|\text{GHZ}_{ijk}\rangle\langle\text{GHZ}_{ijk'}|=I_4\delta_{kk'}$.
\end{enumerate}
\end{lemma}

\noindent \p For the sake of brevity, we again just give the detailed proof of item \ref{000}. It is obvious that when $i=i'$, the conclusion is true according to Lemma \ref{ghz2}. For $i\neq i'$, we have
\begin{eqnarray*}
&\hspace{-5.2cm}\sum\limits^{1}_{j,k=0}Tr_{1}|\text{GHZ}_{ijk}\rangle\langle\text{GHZ}_{\bar{i}jk}|\\
&=\frac{1}{2}\sum\limits_{j,k=0}^{1} (-1)^{\alpha_{ijk}+\alpha_{\bar{i}jk}}  [
(-1)^{\omega(ijk)}|jk\rangle\langle\bar{j}\bar{k}|
+(-1)^{\omega(\bar{i}jk)}|\bar{j}\bar{k}\rangle\langle jk|]\\
&\hspace{-0.22cm}=\frac{1}{2}\sum\limits_{j=0}^1 \{(-1)^{\alpha_{ijj}+\alpha_{\bar{i}jj}}  [
(-1)^{\omega(ijj)}|jj\rangle\langle\bar{j}\bar{j}|
+(-1)^{\omega(\bar{i}jj)}|\bar{j}\bar{j}\rangle\langle jj|]\\
&+(-1)^{\alpha_{ij\bar{j}}+\alpha_{\bar{i}j\bar{j}}}  [
(-1)^{\omega(ij\bar{j})}|j\bar{j}\rangle\langle\bar{j}j|
+(-1)^{\omega(\bar{i}j\bar{j})}|\bar{j}j\rangle\langle j\bar{j}|]\}\\
&\hspace{-9.78cm}=0.
\end{eqnarray*}
\qed

\section{The Proof of Theorem \ref{02}}
\label{sec:AppA}

\noindent \p
Set
\begin{eqnarray}
|\Phi\rangle=\sum\limits^{1}_{i,j,k=0}|\Phi_{ijk}\rangle.
\end{eqnarray}
Denote $C=\{1,2,3\}$ and $Q_{t}=\{1+3t,2+3t,3+3t\}$, $1\leq t\leq m$. For any $S\subseteq\{1,2,\ldots,3+3m\}$, $|S|=3$ and $S^c=\{1,2,\ldots,3+3m\}\backslash S$, the reduced state of subsystem $S$ is given by
\begin{eqnarray}
\rho_S=\sum\limits^{1}_{i,j,k=0}\sum\limits^{1}_{i',j',k'=0}Tr_{S^c}
|\Phi_{ijk}\rangle\langle\Phi_{i'j'k'}|.
\end{eqnarray}
In particular, $\rho_S$ has the following form,
\begin{eqnarray}\label{01}
\rho_S=\sum\limits^{1}_{i,j,k=0}\sum\limits^{1}_{i',j',k'=0}
Tr_{S^c}|\Phi_{ijk}\rangle\langle\Phi_{i'j'k'}|\delta_{ii'}\delta_{jj'}\delta_{kk'}.
\end{eqnarray}

We consider the following cases:

Case 1. $|S\cap C|=3$. The $\rho_S$ has the particular form (\ref{01}) obviously by Lemma \ref{ghz}, and the conclusion holds.

Case 2. $|S\cap C|=2$, $|S\cap Q_x|=1$, $x\in \{1,2,\ldots,m\}$.

If $m\geq 2$, $\rho_S$ has the form (\ref{01}). Since GHZ states are 1-uniform, the conclusion holds by Construction \ref{con1}.

If $m=1$, without of generality, suppose $S\cap C=\{1,2\}$, we consider the following two cases:

(1) When $S\cap Q_1=\{4\}$, by Lemma \ref{ghz1} we have
\begin{eqnarray*}
&\rho_S=\sum\limits^{1}_{i,i',j,j',k=0}|ij\rangle \langle i'j'|\otimes Tr_{23}|\text{GHZ}_{ijk}\rangle\langle\text{GHZ}_{i'j'k}|\\
&\hspace{-0.1cm}=\sum\limits^{1}_{i,i',j,k=0}|ij\rangle \langle i'j|\otimes Tr_{23}|\text{GHZ}_{ijk}\rangle\langle\text{GHZ}_{i'jk}|\\
&\hspace{-3.5cm}=\sum\limits^{1}_{i,i',j=0}|ij\rangle \langle i'j|\otimes I_2\delta_{ii'}\\
&\hspace{-6.8cm}=I_8.
\end{eqnarray*}
\noindent It's similar to the case of  $S\cap Q_1=\{5\}$.

(2) When $S\cap Q_1=\{6\}$, by Lemma \ref{ghz3} we have
\begin{eqnarray*}
&\rho_S=\sum\limits^{1}_{i,i',j,j'=0}|ij\rangle \langle i'j'|\otimes \sum\limits^{1}_{k=0} Tr_{12}|\text{GHZ}_{ijk}\rangle\langle\text{GHZ}_{i'j'k}|\\
&\hspace{-3cm}=\sum\limits^{1}_{i,i',j,j'=0}|ij\rangle \langle i'j'|\otimes I_2\delta_{ii'}\delta_{jj'}\\
&\hspace{-7.3cm}=I_8.
\end{eqnarray*}

Case 3. $|S\cap C|=1$, $|S\cap Q_x|=2$, $x\in \{1,2,\ldots,m\}$.

If $m\geq 2$, $\rho_S$ is given by (\ref{01}).  The conclusion holds by Lemma \ref{ghz2}.

If $m=1$, without of generality, suppose $S\cap C=\{1\}$.  We consider the following two cases:

(1) When $S\cap Q_1=\{4,5\}$, the conclusion holds by Lemma \ref{ghz5}. It is the same for the case of $S\cap Q_1=\{4,6\}$.

(2) When $S\cap Q_1=\{5,6\}$, by Lemma \ref{ghz4} we also get the result.

Case 4.  $|S\cap C|=1$, $|S\cap Q_{x_1}|=1$, $|S\cap Q_{x_2}|=1$, where $x_1\neq x_2$, $x_1,x_2\in \{1,2,\ldots,m\}$. The conclusion establishes for $m\geq 3$ by Lemma \ref{ghz}.

Case 5. $|S\cap C|=0$. The result evidently establishes.
\qed
\section{The properties of Bell states}
\label{sec:AppG}
 For clarity, let us suppose that $x_1,x_2,y_1,y_2,x'_1,x'_2,y'_1,y'_2\in \{0,1\}$, $x=x_1+x_2$ and $y=y_1+y_2$ in the Lemmas \ref{bell1}-\ref{bell3}, in which the summation is taken modulo $2$. Since $\langle\phi_{xy}|\phi_{x'y'}\rangle=\delta_{xx'}\delta_{yy'}$ for  $x,y,x',y'\in\{0,1\}$, the following relations can be straightforwardly proved.
\begin{lemma}\label{bell1}
\begin{enumerate}
\item $\langle\phi_{x_1+x_2,y_1+y_2}|\phi_{x'_1+x_2,y_1+y_2}\rangle=\delta_{x_1x'_1}$
and $\langle\phi_{x_1+x_2,y_1+y_2}|\phi_{x_1+x'_2,y_1+y_2}\rangle=\delta_{x_2x'_2};$
\item $\langle\phi_{x_1+x_2,y_1+y_2}|\phi_{x_1+x_2,y'_1+y_2}\rangle=\delta_{y_1y'_1}$
and $\langle\phi_{x_1+x_2,y_1+y_2}|\phi_{x_1+x_2,y_1+y'_2}\rangle=\delta_{y_2y'_2}$.
\end{enumerate}
\end{lemma}

\begin{lemma}\label{bell2}
\begin{enumerate}
\item $\sum\limits^{1}_{x_2=0}Tr_{2}|\phi_{x_1+x_2,y_1+y_2}\rangle\langle\phi_{x'_1+x_2,y_1+y_2}|
    =I_2\delta_{x_1x'_1}$ and\\
 $\sum\limits^{1}_{x_1=0}Tr_{2}|\phi_{x_1+x_2,y_1+y_2}\rangle\langle\phi_{x_1+x'_2,y_1+y_2}|
 =I_2\delta_{x_2x'_2};$
\item
$\sum\limits^{1}_{y_i=0}Tr_{1}|\phi_{x_1+x_2,y_1+y_2}\rangle\langle\phi_{x'_1+x_2,y_1+y_2}|
=I_2\delta_{x_1x'_1}$\label{003}
and \\ $\sum\limits^{1}_{y_i=0}Tr_{1}|\phi_{x_1+x_2,y_1+y_2}\rangle\langle\phi_{x_1+x'_2,y_1+y_2}|=I_2\delta_{x_2x'_2}$,  $i\in \{1,2\}$.
\end{enumerate}
\end{lemma}
\noindent \p For the first equality in item \ref{000}, we have
\begin{eqnarray*}
&\hspace{-7.8cm}\sum\limits^{1}_{x_2=0}Tr_{2}|\phi_{x_1+x_2,y_1+y_2}
\rangle\langle\phi_{x'_1+x_2,y_1+y_2}|\\
&\hspace{-3.9cm}=\frac{1}{2}\sum\limits^{1}_{x_2=0}\{(-1)^{x_1+x'_1}|x_1+x_2\rangle\langle x'_1+x_2|+|\overline{x_1+x_2}\rangle\langle \overline{x'_1+x_2}|\}\\
&=\frac{1}{2}\{(-1)^{x_1+x'_1}|0\rangle\langle x'_1+x_1|+|1\rangle\langle \overline{x'_1+x_1}|+
(-1)^{x_1+x'_1}|1\rangle\langle \overline{x'_1+x_1}|+|0\rangle\langle x'_1+x_1|\}\\
&\hspace{-12.6cm}=I_2\delta_{x_1x'_1}.
\end{eqnarray*}
\noindent For the first equality in item \ref{003}, let us assume $i=1$. Then we have
\begin{eqnarray*}
&\hspace{-10cm}\sum\limits^{1}_{y_1=0}Tr_{1}|\phi_{x_1+x_2,y_1+y_2}
\rangle\langle\phi_{x'_1+x_2,y_1+y_2}|\\
&\hspace{-0.3cm}=\frac{1}{2}\sum\limits^{1}_{y_1=0}
Tr_{1}\{(-1)^{x_1+x'_1}|x_1+x_2,y_1+y_2\rangle\langle x'_1+x_2,y_1+y_2|+|\overline{x_1+x_2}, \overline{y_1+y_2}\rangle\langle \overline{x'_1+x_2},\overline{y_1+y_2}|\\
&\hspace{0.3cm}+(-1)^{x_1+x_2}|x_1+x_2,y_1+y_2\rangle\langle \overline{x'_1+x_2},\overline{y_1+y_2}|+(-1)^{x'_1+x_2}|\overline{x_1+x_2}, \overline{y_1+y_2}\rangle\langle x'_1+x_2,y_1+y_2|\}.
\end{eqnarray*}

If $x_1= x'_1$, the above equation can be written as
\begin{eqnarray*}
\frac{1}{2}\sum\limits^{1}_{y_1=0}|y_1+y_2\rangle\langle y_1+y_2|+|\overline{y_1+y_2}\rangle\langle \overline{y_1+y_2}|=I_2.
\end{eqnarray*}
\noindent Otherwise for any $y_2\in \{0,1\}$, it can be always written as
\begin{eqnarray*}
&\hspace{-2.1cm}\frac{1}{2}\sum\limits^{1}_{y_1=0}\{(-1)^{x_1+x_2}|y_1+y_2\rangle\langle \overline{y_1+y_2}|+(-1)^{\overline{x_1}+x_2}| \overline{y_1+y_2}\rangle\langle y_1+y_2|\}\\
&=\frac{1}{2}\{(-1)^{x_1+x_2}|0\rangle\langle 1|+(-1)^{\overline{x_1}+x_2}| 1\rangle\langle 0|+(-1)^{x_1+x_2}|1\rangle\langle 0|+(-1)^{\overline{x_1}+x_2}| 0\rangle\langle 1|\}=0.
\end{eqnarray*}

Similar proofs hold for the other cases.
\qed

In addition, since $\sum\limits^{1}_{x,y=0}|\phi_{xy}\rangle\langle\phi_{xy}|=I_4$, we have the following result.

\begin{lemma}\mbox{}\hspace{-8pt}\label{bell3}
$\sum\limits^{1}_{x_i,y_j=0}|\phi_{x_1+x_2,y_1+y_2}\rangle\langle\phi_{x_1+x_2,y_1+y_2}|=I_4$ ~ for any $i,j\in \{1,2\}$.
\end{lemma}
\section{The Proof of Theorem \ref{03}}
\label{sec:AppB}

\noindent \p
Set
\begin{eqnarray}
|\Phi\rangle=\sum\limits^{1}_{i,j,k,f,g=0}|\Phi_{ijkfg}\rangle.
\end{eqnarray}
Let $C=\{1,2,3,4,5\}$, $Q_1=\{6,7\}$, $Q_2=\{8,9\}$, $Q_3=\{10,11\}$ and $Q_{3+t}=\{9+3t,10+3t,11+3t\}$, $1\leq t\leq m$. For any $S\subseteq\{1,2,\ldots,11+3m\}$, $|S|=3$ and $S^c=\{1,2,\ldots,11+3m\}\backslash S$, the reduced state of subsystem $S$ is given by
\begin{eqnarray}
\rho_S=\sum\limits^{1}_{i,j,k,f,g=0}\sum\limits^{1}_{i',j',k',f',g'=0}
Tr_{S^c}|\Phi_{ijkfg}\rangle\langle\Phi_{i'j'k'f'g'}|.
\end{eqnarray}
Denote
\begin{eqnarray}\label{00}
\rho^*_S=\sum\limits^{1}_{i,j,k,f,g=0}\sum\limits^{1}_{i',j',k',f',g'=0}
Tr_{S^c}|\Phi_{ijkfg}\rangle\langle\Phi_{i'j'k'f'g'}|\delta_{ii'}
\delta_{jj'}\delta_{kk'}\delta_{ff'}\delta_{gg'}.
\end{eqnarray}

We consider the following cases:

Case 1. $|S\cap C|=3$. By Lemma \ref{ghz} and Lemma \ref{bell1}, we have $\rho_S=\rho^*_S$. Following with Construction \ref{con1} the conclusion holds.

Case 2.  $|S\cap C|=2$, $|S\cap Q_x|=1$, $x\in \{1,2,\ldots,3+m\}$. It is sufficient to consider the cases of $x=4$ or $x<4$.

If $x<4$, $\rho_S=\rho^*_S$ from Lemma \ref{bell1}. Then the final conclusion establishes since Bell states are 1-uniform except for the cases of $S\cap C=\{1,3\}$ and $|S\cap Q_1|=1$. Hence, we consider the following two cases.

(1) When $S\cap C=\{1,3\}$ and $S\cap Q_1=\{6\}$, by Lemma \ref{bell1} and Lemma \ref{bell2}, we have
\begin{eqnarray*}
&\rho_S=\sum\limits^{1}_{j,f,g=0}\sum\limits^{1}_{i,i',k,k'=0}|ik\rangle \langle i' k'|\otimes Tr_{2}|\phi_{f+i,g+j}\rangle\langle\phi_{f+i',g+j}|\langle\phi_{g+j,i'+k'} |\phi_{g+j,i+k}\rangle\\
&\hspace{-5.7cm}=\sum\limits^{1}_{j,g=0}\sum\limits^{1}_{i,i',k,k'=0}|ik\rangle \langle i' k'|\otimes I_2\delta_{ii'}\delta_{kk'}\\
&\hspace{-10.7cm}=4I_8.
\end{eqnarray*}

(2) When $S\cap C=\{1,3\}$ and $S\cap Q_1=\{7\}$, by Lemma \ref{bell1} and Lemma \ref{bell2}, we have
\begin{eqnarray*}
&\rho_S=\sum\limits^{1}_{j,f,g=0}~\sum\limits^{1}_{i,i',k,k'=0}|ik\rangle \langle i' k'|\otimes Tr_{1}|\phi_{f+i,g+j}\rangle\langle\phi_{f+i',g+j}|\langle\phi_{g+k',f+i'} |\phi_{g+k,f+i}\rangle\\
&\hspace{-5.8cm}=\sum\limits^{1}_{f,g=0}~\sum\limits^{1}_{i,i',k,k'=0}|ik\rangle \langle i' k'|\otimes I_2\delta_{ii'}\delta_{kk'}\\
&\hspace{-11cm}=4I_8.
\end{eqnarray*}

If $x=4$, we have $\rho_S=\rho^*_S$ from Lemma \ref{bell1}. The final conclusion establishes since GHZ states are 1-uniform except for the case of $S\cap C=\{1,5\}$. Hence, we consider the following cases.

(1) When $S\cap C=\{1,5\}$ and $S\cap Q_4=\{12\}$, by Lemma \ref{bell1} we have
\begin{eqnarray*}
&\rho_S=\sum\limits^{1}_{j,k,f=0}\sum\limits^{1}_{i,i',g,g'=0}|ig\rangle \langle i' g'|\otimes Tr_{23}|\text{GHZ}_{jfg}\rangle\langle\text{GHZ}_{jfg'}|\langle\phi_{f+i,g+j} |\phi_{f+i',g'+j}\rangle\\
&\hspace{-5.7cm}=\sum\limits^{1}_{j,k,f=0}\sum\limits^{1}_{i,i',g,g'=0}|ig\rangle \langle i' g'|\otimes \frac{I_2}{2}\delta_{ii'}\delta_{gg'}\\
&\hspace{-11cm}=4I_8.
\end{eqnarray*}
\noindent It is the same for the case of $S\cap Q_4=\{13\}$.

(2) When $S\cap C=\{1,5\}$ and $S\cap Q_4=\{14\}$, by Lemma \ref{ghz1} and Lemma \ref{bell1}  we have
\begin{eqnarray*}
&\rho_S=\sum\limits^{1}_{j,k,f=0}\sum\limits^{1}_{i,i',g,g'=0}|ig\rangle \langle i' g'|\otimes Tr_{12}|\text{GHZ}_{jfg}\rangle\langle\text{GHZ}_{jfg'}|\langle\phi_{g'+k,f+i'} |\phi_{g+k,f+i}\rangle\\
&\hspace{-5.9cm}=\sum\limits^{1}_{k,f=0}\sum\limits^{1}_{i,i',g,g'=0}|ig\rangle \langle i' g'|\otimes I_2\delta_{ii'}\delta_{gg'}\\
&\hspace{-10.9cm}=4I_8.
\end{eqnarray*}

Case 3. $|S\cap C|=1$. We need to consider the following cases:

If $|S\cap Q_x|=2$, it is sufficient to consider the case of $x=4$ or $x<4$. When $x<4$, we have $\rho_S=\rho^*_S$ for any cases by Lemma \ref{bell1}. Then by Lemma \ref{bell3} the conclusion holds. When $x=4$, $\rho_S=\rho^*_S$ holds for any cases. Then by Lemma \ref{ghz2} the conclusion holds.

If $|S\cap Q_{x_1}|=1$, $|S\cap Q_{x_2}|=1$, $x_1\neq x_2$, $ x_1,x_2\in \{1,2,\ldots,m\}$,
we have $\rho_S=\rho^*_S$ by Lemma \ref{ghz} and Lemma \ref{bell1}. Hence, the final conclusion establishes since the Bell states and GHZ states are 1-uniform except for the cases of $S\cap C=\{3\}$, $|S\cap Q_2|=1$ and $|S\cap Q_3|=1$. So we only need to consider the following cases.

(1) When $S\cap Q_2=\{8\}$, $S\cap Q_3=\{10\}$, by Lemma \ref{bell2} we have
\begin{eqnarray*}
&\rho_S=\sum\limits^{1}_{i,f=0} \sum\limits^{1}_{k,k'=0}|k\rangle \langle k'|\otimes \sum\limits^{1}_{g=0} Tr_{2}|\phi_{g+k,f+i}\rangle\langle\phi_{g+k',f+i}|\otimes \sum\limits^{1}_{j=0} Tr_{2}|\phi_{g+j,i+k}\rangle\langle\phi_{g+j,i+k'}|\\
&\hspace{-4.7cm}=\sum\limits^{1}_{i,f=0} \sum\limits^{1}_{k,k'=0}|k\rangle \langle k'|\otimes \sum\limits^{1}_{g=0} Tr_{2}|\phi_{g+k,f+i}\rangle\langle\phi_{g+k',f+i}|\\
&\hspace{-4.6cm}\otimes Tr_{2}\{|\phi_{0,i+k}\rangle\langle\phi_{0,i+k'}|+|\phi_{1,i+k}\rangle\langle\phi_{1,i+k'}|\}\\
&\hspace{-7.7cm}=\sum\limits^{1}_{i,f=0} \sum\limits^{1}_{k,k'=0}|k\rangle \langle k'|\otimes I_2  \otimes I_2 \delta_{kk'}\\
&\hspace{-12.2cm}=4I_8.
\end{eqnarray*}

(2) When $S\cap Q_2=\{8\}$, $S\cap Q_3=\{11\}$, by Lemma \ref{bell2} we have
\begin{eqnarray*}
&\rho_S=\sum\limits^{1}_{i,j,f=0} \sum\limits^{1}_{k,k'=0}|k\rangle \langle k'|\otimes \sum\limits^{1}_{g=0}Tr_{2}|\phi_{g+k,f+i}\rangle\langle\phi_{g+k',f+i}|\otimes Tr_{1}|\phi_{g+j,i+k}\rangle\langle\phi_{g+j,i+k'}|\\
&\hspace{-3.4cm}=\sum\limits^{1}_{i,j,f=0} \sum\limits^{1}_{k,k'=0}|k\rangle \langle k'|\otimes I_2 \delta_{kk'}\otimes Tr_{1}|\phi_{g+j,i+k}\rangle\langle\phi_{g+j,i+k'}|\\
&\hspace{-11.8cm}=4I_8.
\end{eqnarray*}

(3) When $S\cap Q_2=\{9\}$, $S\cap Q_3=\{10\}$, by Lemma \ref{bell2} we have
\begin{eqnarray*}
&\rho_S=\sum\limits^{1}_{i,j,g=0} \sum\limits^{1}_{k,k'=0}|k\rangle \langle k'|\otimes \sum\limits^{1}_{f=0}Tr_{1}|\phi_{g+k,f+i}\rangle\langle\phi_{g+k',f+i}|\otimes Tr_{2}|\phi_{g+j,i+k}\rangle\langle\phi_{g+j,i+k'}|\\
&\hspace{-3.4cm}=\sum\limits^{1}_{i,j,g=0} \sum\limits^{1}_{k,k'=0}|k\rangle \langle k'|\otimes I_2 \delta_{kk'}\otimes Tr_{2}|\phi_{g+j,i+k}\rangle\langle\phi_{g+j,i+k'}|\\
&\hspace{-11.8cm}=4I_8.
\end{eqnarray*}
\noindent It is the same for the case of $S\cap Q_3=\{11\}$.

Case 4.  $|S\cap C|=0$. The conclusion obviously holds. \qed

\section{The Proof of Construction \ref{qud}}
\label{sec:AppD}

\noindent \p
Set
\begin{eqnarray}
|\Phi\rangle=\sum\limits_{i,j,k\in[d]}|\Phi_{ijk}\rangle.
\end{eqnarray}
Let $C=\{1,2,3,4\}$, $Q_{t}=\{2+3t,3+3t,4+3t\}$, $1\leq t\leq m$. For any $S\subseteq\{1,2,\ldots,4+3m\}$, $|S|=3$ and $S^c=\{1,2,\ldots,4+3m\}\backslash S$, the reduced state of subsystem $S$ is given by
\begin{eqnarray}
\rho_S=\sum\limits_{i,j,k\in[d]}\sum\limits_{i',j',k'\in[d]}Tr_{S^c}|
\Phi_{ijk}\rangle\langle\Phi_{i'j'k'}|.
\end{eqnarray}
Denote
\begin{eqnarray}\label{11}
\rho^*_S=\sum\limits_{i,j,k\in[d]}\sum\limits_{i',j',k'\in[d]}
Tr_{S^c}|\Phi_{ijk}\rangle\langle\Phi_{i'j'k'}|\delta_{ii'}\delta_{jj'}\delta_{kk'}.
\end{eqnarray}
We need to consider the following cases:

Case 1. $|S\cap C|=3$. We obviously have $\rho_S=\rho^*_S$ by Lemma \ref{phi}. Combining with the Construction \ref{04} the conclusion holds.

Case 2. $|S\cap C|=2$, $|S\cap Q_x|=1$ and $x\in \{1,2,\ldots,m\}$. If $m\geq 2$, the conclusion holds by Construction \ref{04} and Lemma \ref{phi}.
Otherwise, when $S\cap C=\{1,2\}$, by Lemma \ref{phi} we have
\begin{eqnarray*}
&\rho_{125}
=\sum\limits_{i,j,k,i',j',k'\in [d]}Tr_{3,4}|i,k,i+j+k,i+\alpha_2j+\alpha_2^2k\rangle \langle i',k',i'+j'+k',i'+\alpha_2j'+\alpha_2^2k'|\\
&\hspace{-9cm}\otimes Tr_{2,3} |\psi_{ijk}\rangle\langle \psi_{i'j'k'}|\\
&\hspace{0.2cm}=\frac{1}{d}\sum\limits_{i,j,k,i',j',k'\in [d]}|i,k\rangle\langle i',k'|\langle i'+j'+k'|i+j+k\rangle \langle i'+\alpha_2j'+\alpha_2^2k'|i+\alpha_2j+\alpha_2^2k\rangle\\
&\otimes\sum\limits_{l\in [d]} \omega^{il-i'l'}|l+j\rangle\langle l'+j'|\langle l'+\alpha_2j'+\beta k'|l+\alpha_2j+\beta k\rangle\langle l'|l\rangle\delta_{i'i}\delta_{j'j}\delta_{k'k}\\
&\hspace{-12.9cm}=\mathbb{I}_{d^3}.
\end{eqnarray*}
The second equality in the above derivation is true since from the following set of equations,
\begin{equation}
    \left\{
          \begin{array}{ll}
i'+j'+k'=i+j+k,\\
i'+\alpha_2j'+\alpha_2^2k'=i+\alpha_2j+\alpha_2^2k,\\
l'+\alpha_2j'+\beta k'=l+\alpha_2j+\beta k,\\
l'=l,\\
          \end{array}\label{30}
       \right.
\end{equation}
\noindent we can get
\begin{equation}
    \left\{
          \begin{array}{ll}
(\alpha_2-1)(j'-j)+(\alpha_2^2-1)(k'-k)=0,\\
\alpha_2(j'-j)+\beta (k'-k)=0.\\
          \end{array}
       \right.
\end{equation}
Because $\beta \neq \alpha_2^2+\alpha_2$, there is only one solution $i=i',j=j',k=k'$ and $l=l'$ for Eq. (\ref{30}).
\begin{eqnarray*}
&\rho_{126}
=\sum\limits_{i,j,k,i',j',k'\in [d]}Tr_{3,4}|i,k,i+j+k,i+\alpha_2j+\alpha_2^2k\rangle \langle i',k',i'+j'+k',i'+\alpha_2j'+\alpha_2^2k'| \\
 &\hspace{-9cm}\otimes Tr_{1,3} |\psi_{ijk}\rangle\langle \psi_{i'j'k'}|\\
 &\hspace{0.2cm}= \frac{1}{d}\sum\limits_{i,j,k,i',j',k'\in [d]}|i,k\rangle\langle i',k'|\langle i'+j'+k'|i+j+k\rangle \langle i'+\alpha_2j'+\alpha_2^2k'|i+\alpha_2j+\alpha_2^2k\rangle\\
 &\otimes\sum\limits_{l\in [d]} \omega^{il-i'l'}|l+\alpha_2j+\beta k\rangle \langle l'+\alpha_2j'+\beta k'  |\langle l'+j'| l+j \rangle \langle l'|l\rangle\delta_{i'i}\delta_{j'j}\delta_{k'k}\\
 &\hspace{-12.9cm}=\mathbb{I}_{d^3},
 \end{eqnarray*}
\noindent where the second equality is true since $\alpha_2^2\neq 1$. There is only one solution $i=i',j=j',k=k'$ and $l=l'$ for the following set of equations,
\begin{equation}
    \left\{
          \begin{array}{ll}
i'+j'+k'=i+j+k,\\
i'+\alpha_2j'+\alpha_2^2k'=i+\alpha_2j+\alpha_2^2k,\\
l'+j'=l+j,\\
l'=l.\\
          \end{array}
       \right.
\end{equation}
 \begin{eqnarray*}
 &\rho_{127}
=\sum\limits_{i,j,k,i',j',k'\in [d]}Tr_{3,4}|i,k,i+j+k,i+\alpha_2j+\alpha_2^2k\rangle \langle i',k',i'+j'+k',i'+\alpha_2j'+\alpha_2^2k'|\\
 &\hspace{-9cm}\otimes Tr_{1,2} |\psi_{ijk}\rangle\langle \psi_{i'j'k'}|\\
 &\hspace{0.2cm}= \frac{1}{d}\sum\limits_{i,j,k,i',j',k'\in [d]}|i,k\rangle\langle i',k'|\langle i'+j'+k'|i+j+k\rangle \langle i'+\alpha_2j'+\alpha_2^2k'|i+\alpha_2j+\alpha_2^2k\rangle\\
 &\otimes\sum\limits_{l\in [d]} \omega^{il-i'l'} |l'\rangle\langle l| \langle l'+j'| l+j \rangle\langle l'+\alpha_2j'+\beta k'|l+\alpha_2j+\beta k\rangle  \delta_{l'l}\delta_{i'i}\delta_{j'j}\delta_{k'k}\\
 &\hspace{-12.9cm}=\mathbb{I}_{d^3},
  \end{eqnarray*}

\noindent where the second equality is true since $\beta \neq \alpha_2^2-1$. There is only one solution $i=i',j=j',k=k'$ and $l=l'$ for the following set of equations,
\begin{equation}
    \left\{
          \begin{array}{ll}
i'+j'+k'=i+j+k,\\
i'+\alpha_2j'+\alpha_2^2k'=i+\alpha_2j+\alpha_2^2k,\\
l'+j'=l+j,\\
l'+\alpha_2j'+\beta k'=l+\alpha_2j+\beta k.\\
          \end{array}
       \right.
\end{equation}
It is easy to check that the other cases of  $S\cap C=\{1,3\},\{1,4\},\{2,3\},\{2,4\}$ and $\{3,4\}$  are also true since $\beta \neq \alpha_2, \alpha_2^2, \alpha_2-1$, $\alpha_2^2-\alpha_2$ and $\alpha_2\neq 1$.

Case 3. $|S\cap C|=1, |S\cap Q_x|=2, x\in \{1,2,\ldots,m\}$. Since any three columns in the first four columns of $|\Phi\rangle$ form an OA$(d^3,3,d,3)$, we have $\rho_S=\rho^*_S$. According to Lemma \ref{phi1}, the conclusion holds except for the case of $S\cap C= \{3\}$ or \{4\}. In the following, we show that any triple $(x,y,z)$ appears exactly $d$ times in the columns 123, 124 and 134 of the array $(i+\alpha_sj+\alpha_s^2k,l+j,l+\alpha_2j+\beta k,l)$ for $i,j,k,l\in[d]$, where $s=1,2$ and $\alpha_1=1$. Concerning the columns 123, we have
\begin{equation}
    \left\{
          \begin{array}{ll}
i+\alpha_sj+\alpha_s^2k=x,\\
l+j=y,\\
l+\alpha_2j+\beta k=z.\\
          \end{array}
       \right.
\end{equation}
It is easy to check that  for any $i\in [d]$ there exists one solution: $j=\alpha_s^{-1}[\beta -\alpha_s(\alpha_2-1)]^{-1}[\beta (x-i)-\alpha_s^2(z-y)]$, $k=\alpha_s^{-1}[\beta -\alpha_s(\alpha_2-1)]^{-1}[\alpha_s(z-y)-(\alpha_2-1)(x-i)]$ and $l=y-\alpha_s^{-1}[\beta -\alpha_s(\alpha_2-1)]^{-1}[\beta (x-i)-\alpha_s^2(z-y)]$, where $\beta \neq\alpha_s(\alpha_2-1)$.

With respect to the columns 124, we have
\begin{equation}
    \left\{
          \begin{array}{ll}
i+\alpha_sj+\alpha_s^2k=x,\\
l+j=y,\\
l=z.\\
          \end{array}
       \right.
\end{equation}
For any $i\in [d]$, there exists one solution that $j=y-z$, $k=\alpha_s^{-2}[x-i-\alpha_s(y-z)]$ and $l=z$.

For the columns 134, we have
\begin{equation}
    \left\{
          \begin{array}{ll}
i+\alpha_sj+\alpha_s^2k=x,\\
l+\alpha_2j+\beta k=y,\\
l=z.\\
          \end{array}
       \right.
\end{equation}
For any $i\in [d]$, the solution is $j=\alpha_s^{-1}(\beta -\alpha_2\alpha_s)^{-1}[\beta (x-i)-\alpha_s^2(y-z)]$, $k=\alpha_s^{-1}(\beta -\alpha_2\alpha_s)^{-1}[\alpha_s(y-z)-\alpha_2(x-i)]$ and $l=z$,  where $\beta \neq\alpha_2\alpha_s$.

Case 4. $|S\cap C|=1, |S\cap Q_{x_1}|=1, |S\cap Q_{x_2}|=1, x_1\neq x_2$, $x_1,x_2\in \{1,2,\dots,m\}$. By Lemma \ref{phi}, the conclusion holds.

Case 5. $|S\cap C|=0$. The conclusion clearly establishes. \qed

\section{The known results on quantum orthogonal arrays of strength 2}
\label{sec:AppE}

In Ref. \cite{Goyeneche1}, some results on quantum orthogonal arrays of strength 2 with levels of any prime number are given.
\begin{lemma}\mbox{}\hspace{-8pt}\label{oa0} \ {\rm (\cite{Goyeneche1})}
Let $|\Phi^{\pm}\rangle=(|00\rangle\pm|11\rangle)/\sqrt{2}$ and
$|\Psi^{\pm}\rangle=(|01\rangle\pm|10\rangle)/\sqrt{2}$.
\begin{enumerate}
\item The arrangement
\begin{equation}\left(
\begin{array}{cccc}
|0\rangle & |0\rangle  &|\Phi^+\rangle^{\otimes m}\\
|0\rangle & |1\rangle  &|\Psi^+\rangle^{\otimes m}\\
|1\rangle & |0\rangle  &|\Psi^-\rangle^{\otimes m}\\
|1\rangle & |1\rangle &|\Phi^-\rangle^{\otimes m}
\end{array}\right)
\end{equation}
is a QOA($4,2+2m,2,2$) with $m\geq 2$.
\item The arrangement
\begin{equation}\left(
\begin{array}{cccc}
|0\rangle & |0\rangle & |0\rangle &|\Phi^+\rangle^{\otimes m}\\
|0\rangle & |1\rangle & |1\rangle &|\Psi^+\rangle^{\otimes m}\\
|1\rangle & |0\rangle & |1\rangle &|\Psi^-\rangle^{\otimes m}\\
|1\rangle & |1\rangle & |0\rangle &|\Phi^-\rangle^{\otimes m}
\end{array}\right)
\end{equation}
is a QOA($4,3+2m,2,2$) with $m\geq 1$.
\item For any odd prime number $d$ and $m\geq1$, let
$|\varphi^1_{ij}\rangle=|i,j,i+j \rangle |\phi_{ij}\rangle^{\otimes m}$ and
$|\varphi^2_{ij}\rangle=|i,j,i+j,i+2j \rangle |\phi_{ij}\rangle^{\otimes m}$,
where $|\phi_{ij}\rangle=\frac{1}{\sqrt{d}}\sum\limits_{l\in[d]}\omega^{il}|l+j,l\rangle$ with $\omega=e^{\frac{2\pi\sqrt{-1}}{d}}$. Then $\{|\varphi^1_{ij}\rangle:i,j\in [d]\}$ forms a QOA$(d^2,3+2m,d,2)$ and $\{|\varphi^2_{ij}\rangle:i,j\in [d]\}$ forms a QOA$(d^2,4+2m,d,2)$.
\end{enumerate}
\end{lemma}




%
%


\section*{References}
\begin{thebibliography}{}
\bibitem{Benenti}
Benenti, G.,  Casati, G.,  Rossini, D.,  Strini, G.:
{\em Principles of quantum computation and information};
World Scientific Publishing, Singapore, (2019)
\bibitem{Chuang}
 Nielsen, M.A.,  Chuang, I.L.: {\em Quantum computation and quantum information}; Cambridge University Press, Cambridge, United Kingdom, (2000)
\bibitem{Jozsa}
 Jozsa, R.,  Linden, N.: On the role of entanglement in quantum computational speed-up. {\em Proc. R. Soc. A},  {\bf 459}, 2011-2032 (2003)
\bibitem{BennettC2}  Bennett, C. H.,  Brassard, G.,  Cr\'{e}peau, C.,  Jozsa, R.,  Peres, A.,  Wootters, W.K.:  Teleporting an unknown quantum state via dual classical and einstein-podolsky-rosen channels. {\em Phys. Rev. Lett.}, {\bf 70}, 1895 (1993)

\bibitem{BennettC1}  Bennett, C.H.: Quantum cryptography using any two nonorthogonal states. {\em Phy. Rev. Lett.},  {\bf 68}, 3121 (1992)
\bibitem{Lo}
 Lo, H.K.,  Curty, M.,  Qi, B.: Measurement-device-independent quantum key distribution. {\em Phys. Rev. Lett.},  {\bf 108}, 130503 (2012)
\bibitem{Goyeneche0}
 Goyeneche, D.,  \.{Z}yczkowski, K.: Genuinely multipartite entangled states and orthogonal arrays. {\em Phys. Rev. A},  {\bf 90}, 022316 (2014)
\bibitem{Latorre}
 Latorre, J.I.,  Sierra, G.: Holographic codes.
arXiv:1502.06618 (2015).
\bibitem{Zhang}
 Zhang, J.,  Adesso, G.,  Xie, C., Peng, K.: Quantum teamwork for unconditional multiparty communication with Gaussian states.
{\em Phys. Rev. Lett.,}  {\bf 103}, 070501 (2009)
\bibitem{Facchi1}
 Facchi, P.: Multipartite entanglement in qubit systems. {\em Rend. Lincei Mat. Appl.}, {\bf 20}, 25-67 (2009)
\bibitem{Facchi2}
 Facchi, P.,  Florio, G.,  Parisi, G.,  Pascazio, S.: Maximally multipartite entangled states. {\em Phys. Rev. A},   {\bf 77},  060304 (2008)
\bibitem{Helwig}
 Helwig, W.,  Cui, W., Latorre, J.I.,  Riera, A., Lo, H.K.:  Absolute maximal entanglement and quantum secret sharing. {\em Phys. Rev. A}, {\bf 86},  052335 (2012)
\bibitem{Higuchi}
 Higuchi, A.,  Sudbery, A.: How entangled can two couples get?  {\em Phys. Lett. A},  {\bf 273}, 213-217 (2000)
\bibitem{Li}
 Li, M.S.,  Wang, Y.L.: $k$-uniform quantum states arising from orthogonal arrays. {\em Phy. Rev.  A}, {\bf 99}, 042332 (2019)
\bibitem{Pang1}
 Pang, S.Q.,  Zhang, X., Lin, X.,   Zhang, Q.J.: Two and three-uniform states from irredundant orthogonal arrays. {\em npj Quantum Inf.},  {\bf 5}, 1-10 (2019)
\bibitem{Rains}
 Rains, E.M.: Nonbinary quantum codes. {\em IEEE Trans. Inf. Theory}, {\bf 45}, 1827-1832 (1999)

\bibitem{Grassl}
Grassl M.: Code tables: bounds on the parameters of various types of codes. http://www.codetables.\\de/ (accessed 08 November 2022)
\bibitem{Rather}
 Rather, S.A.,  Burchardt, A.,   Bruzda, W., Rajchel-Mieldzio$\acute{c}$, G.,  Lakshminarayan, A., $\dot{Z}$yczkowski, K.: Thirty-six entangled officers of Euler: Quantum solution to a classically impossible problem. {\em Phys. Rev. Lett}, {\bf 128}, 080507 (2022)
\bibitem{Scott}
 Scott, A.J.: Multipartite entanglement, quantum-error-correcting codes, and entangling power of quantum evolutions.
{\em Phy. Rev. A}, {\bf 69}, 052330 (2004)
\bibitem{Zang1}
 Zang, Y.,  Chen, G.Z.,  Chen, K.J.,  Tian, Z.: Further results on 2-uniform states arising from irredundant orthogonal arrays. {\em Adv. Math. Commun.},  {\bf 16}, 231-247 (2022)

\bibitem{Zang2}
 Zang, Y.,  Facchi, P.,  Tian, Z.: Quantum combinatorial designs and $k$-uniform states. {\em J. Phys. A: Math. Theor.}, {\bf 54}, 505204 (2021)
\bibitem{Huber1}
Huber, F.,   G$\ddot{u}$hne, O.,  Siewert, J.: Absolutely maximally entangled states of seven qubits do not exist.  {\em Phys. Rev.
Lett.},  {\bf 118}, 200502 (2017)
\bibitem{Zang}
 Zang, Y.,  Zuo, H.J., Tian, Z.: 3-uniform states and orthogonal arrays of strength 3. {\em Int. J. Quantum Information},  {\bf 17}, 1950003 (2019)
\bibitem{Huber}
Huber, F.,   Wyderka, N.: Table of absolutely maximally entangled states. http://tp.nt.uni-siegen.de/
+fhuber/ame.html (2020). Accessed 4 May 2020
\bibitem{Chen}
 Chen, G.Z., Zhang, X.T.: Constructions of irredundant orthogonal arrays. {\em Adv. Math. Commun.}. Available online: http://www.doi:10.3934/amc.2021051. (2021)
\bibitem{Pang2}
 Pang, S.Q.,  Zhang, X.,  Du, J., Wang, T.Y.:  Multipartite entanglement states of higher uniformity. {\em J. Phys. A: Math. Theor.}, {\bf 54}, 015305 (2021)
\bibitem{Goyeneche1}
 Goyeneche, D.,  Raissi, Z.,   Martino, S. Di, \.{Z}yczkowski, K.: Entanglement and quantum combinatorial designs. {\em Phys. Rev. A},  {\bf 97}, 062326 (2018)
\bibitem{Raissi}
 Raissi, Z., Teixid$\acute{o}$, A.,  Gogolin, C., Ac$\acute{\imath}$n, A.: Constructions of $k$-uniform and absolutely maximally entangled states
beyond maximum distance codes. {\em Phys. Rev. Research},  {\bf 2}, 033411 (2020)
\bibitem{Seiden}
 Seiden, E., Zemach, R.: On orthogonal arrays. {\em Ann. Math. Stat.},  {\bf 37}, 1355-1370 (1966)
\bibitem{Hedayat}
 Hedayat, A.S., Sloane, N.J.A., Stufken, J.:  {\em Orthogonal array: theory and applications}; Springer-Verlag, (1999)
\bibitem{Arkin}
 Arkin, J.,  Straus, E.G.: Latin $k$-cubes. {\em Fibonacci Quart.,} {\bf 12},  288-292 (1974)
\bibitem{Lidl}
 Lidl, R.,  Niederreiter, H.:  {\em Finite fields} (2nd Edition); Cambridge University Press, (1997)
\bibitem{Bierbrauer}
 Bierbrauer, J.: Construction of orthogonal arrays.  {\em J. Stat. Plann. Infer.},  {\bf 56}, 39-47 (1996)
\bibitem{Mukhopadhyay}
 Mukhopadhyay, A.C.: Construction of some series of orthogonal arrays. {\em Sankhy$\bar{a}$ Ser. B}, {\bf 43}, 81-92 (1981)
\bibitem{Stinson}
 Stinson, D.R.: Ideal ramp schemes and related combinatorial objects. {\em Discrete Math.},
{\bf 341}, 299-307 (2018)

\end{thebibliography}


\end{document}